\documentclass[]{spie}  

\usepackage{amsmath,amsfonts,amssymb}
\usepackage{graphicx}
\usepackage{mathtools}
\usepackage[colorlinks=true, allcolors=blue]{hyperref}

\title{A data-driven approach to sampling matrix selection for compressive sensing}

\author[a]{Elin Farnell}
\author[a]{Henry Kvinge}
\author[b]{John P. Dixon}
\author[b]{Julia R. Dupuis}
\author[a]{Michael Kirby}
\author[a]{Chris Peterson}
\author[b]{Elizabeth C. Schundler}
\author[b]{Christian W. Smith}

\affil[a]{Colorado State University, Department of Mathematics, 1874 Campus Delivery, Fort Collins, CO 80523-1874, USA}
\affil[b]{Physical Sciences Inc., 20 New England Business Center, Andover, MA 01810-1077, USA}

\authorinfo{Further author information (E-mail): (Send correspondence to E.F.)\\E.F.: elinfarnell@gmail.com;
H.K.: henry.kvinge@colostate.edu; M.K.: kirby@math.colostate.edu;\\ C.P.: peterson@math.colostate.edu; J.R.D.: jdupuis@psicorp.com; C.W.S.: cwsmith@psicorp.com;\\ E.C.S.: eschundler@psicorp.com; J.P.D.: jdixon@psicorp.com}

\begin{document} 
\maketitle

\begin{abstract}
Sampling is a fundamental aspect of any implementation of compressive sensing. Typically, the choice of sampling method is guided by the reconstruction basis. However, this approach can be problematic with respect to certain hardware constraints and is not responsive to domain-specific context. We propose a method for defining an order for a sampling basis that is optimal with respect to capturing variance in data, thus allowing for meaningful sensing at any desired level of compression. We focus on the Walsh-Hadamard sampling basis for its relevance to hardware constraints, but our approach applies to any sampling basis of interest. We illustrate the effectiveness of our method on the Physical Sciences Inc. Fabry-P\'{e}rot interferometer sensor multispectral dataset, the Johns Hopkins Applied Physics Lab FTIR-based longwave infrared sensor hyperspectral dataset, and a Colorado State University Swiss Ranger depth image dataset. The spectral datasets consist of simulant experiments, including releases of chemicals such as GAA and SF6. We combine our sampling and reconstruction with the adaptive coherence estimator (ACE) and bulk coherence for chemical detection and we incorporate an algorithmic threshold for ACE values to determine the presence or absence of a chemical. We compare results across sampling methods in this context. We have successful chemical detection at a compression rate of 90\%. For all three datasets, we compare our sampling approach to standard orderings of sampling basis such as random, sequency, and an analog of sequency that we term `frequency.'  In one instance, the peak signal to noise ratio was improved by over 30\% across a test set of depth images.
\end{abstract}


\section{INTRODUCTION}
\label{sec:intro}  

Many promising kinds of imaging currently have limited application because the sensors required are either too expensive, too slow, or too large for a given application. One of the attractive aspects of the compressive sensing (CS) framework is that it allows for alternative sampling routines that can sometimes overcome these limitations. The single pixel camera proposed by Duarte et. al. \cite{duarte2008single} is a good example of this; the use of CS in MRI is another \cite{lustig2007sparse}.

Much of the theory behind CS attempts to understand which sampling techniques maximize accuracy in a reconstructed signal. Foundational work in the field provides reconstruction guarantees based on whether the sampling matrix $S$ (which controls how the signal is sampled) satisfies the restricted isometry property \cite{candes2005decoding}. When sampling is actually done via a product of matrices $SD$ where $D$ is fixed (as often happens in practice when the signal is not sparse in its natural basis), then $S$ can be chosen via its incoherence \cite{donoho2001uncertainty} with $D$. Incoherence itself can (with high probability) be obtained by drawing entries of $S$ randomly from some known distribution. This scheme has been used for many specific applications of CS \cite{takhar2006new,lustig2007sparse,baraniuk2007compressive}. It has also been shown to be flexible enough to give reconstruction guarantees when the specific structure is expected between samples \cite{baraniuk2010model}.

In this paper we take a different approach. Rather than designing a sampling scheme which is guaranteed to be able to reconstruct any signal, we instead choose $S$ based on its ability to capture the maximum amount of variance on a set of training signals (or in other words, a set of signals representative of the signals we expect to later sample and reconstruct). We show that when $S$ is constructed in such a way it performs similarly or better than other popular sampling schemes. Based on these results we propose this method for CS applications where a set of representative example signals is known before imaging begins. 

To simplify the analysis (and also to make the results applicable to the single-pixel-camera framework) we focus on sampling matrices $S$ whose entries consist of $0$'s and $1$'s. Specifically, we study sampling matrices generated from subsets of rows of the shifted Walsh-Hadamard matrix. In order to show the flexibility of our method we present results for reconstruction of both LIDAR and hyperspectral data (the Physical Sciences Inc. Fabry-P\'{e}rot interferometer sensor multispectral dataset, the Johns Hopkins Applied Physics Lab FTIR-based longwave infrared sensor hyperspectral dataset, and the Colorado State University Swiss Ranger depth image dataset)~\cite{cosofret2009airis,broadwater2011primer}. One particular advantage is that since rows of the matrix $S$ are ranked by the amount of variance they capture in the data, $S$ does not need to be recalculated when decreasing the sampling level. Instead we can simply omit the appropriate rows of $S$.

This paper is organized as follows. In Secs.~\ref{sec:cs}-\ref{sec:opt} we review the basics of CS and state the optimization problem that we are solving in order to reconstruct signals. In \ref{sec:RIP} we review the ideas that underlie the standard methods for generating a sampling matrix. We state and describe the metrics that we used to analyze signal reconstruction accuracy in Sec.~\ref{sec:PSNRRMSE}. Since two of the datasets used to measure the performance of our method consist of hyperspectral data, in Sec.~\ref{sec:ACEBCandThresholds} we describe the algorithms that we used to detect signals in this context. In Sec.~\ref{sec:data-driven}, we describe our maximal-variance based method of generating a sampling matrix. In this section we also give performance results for our method on both hyperspectral and LIDAR datasets. Finally, we conclude and suggest directions for future work in Sec.~\ref{sec:conclusion}.

\section{BACKGROUND}
\subsection{Compressive Sensing}
\label{sec:cs}

Compressive sensing (CS) is a collection of methods that, in certain cases, allows for highly accurate reconstruction of signals even when they have been sampled well under the Nyquist-rate~\cite{baraniuk2007compressive,willett2014sparsity}. Framed in terms of linear algebra (as much of this paper is), CS can be equivalently formulated as being a set of methods for solving the ill-posed problem of finding $\tilde{x}$ from 
\begin{equation*}
y = S\tilde{x} \in \mathbb{R}^k
\end{equation*}
when $S$ is a $k \times n$ matrix, $\tilde{x} \in \mathbb{R}^n$, and $k < n$. CS methods used to solve such problems utilize additional assumptions about the signal $\tilde{x}$. These additional assumptions allow one to choose the correct signal $\tilde{x}$ out of the infinite number of solutions $x'$ to $Sx' = y$. For example, it is often the case that $\tilde{x}$ will be sparse (or compressible) in some particular basis. Formally this means that there is some $n \times n$ basis transformation matrix $D$ such that $\tilde{x} = D\tilde{u}$ and $\tilde{u}$ is sparse. 

The matrix $S$ is known as the {\emph{sampling matrix}} and $D$ is known as the {\emph{sparsity-promoting matrix}}. As its name suggests, the sampling matrix defines the way in which samples are collected from a signal. In many CS applications, one has some latitude to choose $S$ up to context-specific constraints, and hence understanding which matrices $S$ will capture the maximum amount of information about a signal from a minimum number of samples is a question of central importance to the CS community. As a consequence, rich theory has been developed to explain the relative performance of different sampling matrices (see Sec.~\ref{sec:RIP} for a brief overview). In this paper we will present a new data-driven approach to the choice of $S$.

\subsection{Reconstruction of signal via optimization}
\label{sec:opt}

Given a sampled signal $y = SD\tilde{u}$, CS theory shows that, provided that $S$ has been well-chosen and $\tilde{u}$ has certain basic characteristics (such as sparsity), a close approximation $x$ to $\tilde{x}$ can be reconstructed by solving an optimization problem. In this paper, the optimization problem that was solved in order to approximate $\tilde{x}$ from the set of measurements $y = S\tilde{x} \in \mathbb{R}^k$ with $k \ll n$ is
\begin{equation} \label{eqn-opt-problem}
    \text{argmin}_{u \in \mathbb{R}^{n}} ||u||_{\ell_1} \quad\quad \text{such that } \quad y = SH^{-1}u.
\end{equation}
Here $H,H^{-1}: \mathbb{R}^n \rightarrow \mathbb{R}^n$ is the Haar wavelet transform and its inverse, respectively. 
Informally, this optimization problem requires that we find a solution $u$ which satisfies $SH^{-1}u = y$ such that $||u||_{\ell_1}$ is minimized. Minimizing the term $||u||_{\ell_1}$ typically leads to a solution which is sparse in the wavelet domain. At an intuitive level, this makes sense as many types of signals (particularly various kinds of images) have been shown to be naturally sparse in wavelet bases. The attentive reader will notice that the $\ell_1$-norm is used rather than the $\ell_0$ pseudo-norm (which measures sparsity directly). This substitution is common practice in CS since the $\ell_1$-norm is convex whereas the $\ell_0$-pseudo norm is not and results in problems which are impractical to solve.

Driven in part by the interest in CS, there are now many methods of solving $\ell_1$-type optimization problems. We choose to use the split Bregman method \cite{GO09} to solve \eqref{eqn-opt-problem} because it provides fast, lightweight reconstruction with reliable convergence properties.

\subsection{Restricted Isometry Property and Incoherence}
\label{sec:RIP}

In this section we review one of the most popular approaches to producing sampling matrices that guarantee good reconstructions of signals. These ideas hinge on a characterization of matrices called the {\emph{restricted isometry property}} (RIP) \cite{candes2008restricted}. Our introduction of this topic follows the outline given by Baraniuk \cite{baraniuk2007compressive}. 

In the unrealistic situation where we know that $\tilde{u}$ is $p$-sparse and we know the exact coefficients in which $\tilde{u}$ is sparse, then we only need to consider the corresponding $p$ columns of $A = SD$ (or for optimization problem \eqref{eqn-opt-problem}, the corresponding $p$ columns of $SH^{-1}$). In particular, for $A$ to have a well-conditioned inverse on all $p$-sparse vectors $z$ following the sparsity pattern of $\tilde{u}$ we need to ensure that for all such $z$ the following inequality is satisfied:
\begin{equation} \label{eqn-RIP}
    (1 - \delta_p) \leq \frac{||Az||_{\ell_2}^2}{||z||_{\ell_s}^2} \leq (1+\delta_p).
\end{equation}
In general of course, while we may have a guess for the natural sparsity of our signal $\tilde{u}$, we rarely know which specific $p$ coefficients of $\tilde{u}$ will be nonzero (and in most real-world situations, $u$ will not be sparse in the sense of having many terms be exactly zero, but instead will be {\emph{compressible}} in the sense of many terms being very close to zero). To generalize to such a situation, $A$ is said to have the $p$-{\emph{restricted isometry property}} provided that for any $p$-sparse vector $z$, \eqref{eqn-RIP} holds. It has been shown that recovery of signals that are either sparse or compressible is guaranteed when $A$ satisfies a specific RIP depending on the assumed sparsity of the signal.

However, even verifying that a matrix satisfies an RIP property is a combinatorially difficult problem. Furthermore, in our approach we can only change $S$, not $H^{-1}$ which is already assumed to be fixed. In order to try to obtain a product, $SH^{-1}$, which guarantees good reconstruction properties, another (related) option is to try to maximize the {\emph{incoherence}} between $S$ and $H^{-1}$. Roughly, this means constructing $S$ such that the columns of $S$ cannot sparsely represent the columns of $H^{-1}$. 

The standard way to either find a matrix $S$ satisfying an RIP property or $S$ such that $S$ and $H^{-1}$ are incoherent is to utilize randomness. It has been shown that if $S$ is drawn randomly from a number of standard distributions, then with high probability $S$ and $H^{-1}$ will be incoherent and $SH^{-1}$ will satisfy the RIP property \cite{candes2006robust,donoho2006compressed,baraniukjohnson}. Of course, the methods above produce $S$ with good properties in general. In some cases however, characteristics of the signal we wish to sample are known sufficiently well that it makes sense to try to adapt $S$ to best capture information from this specific class of signals. Additionally, hardware constraints sometimes make the methods above inaccessible. Presenting a procedure that addresses this setting is the purpose of this paper.

\subsection{Image Quality Metrics}
\label{sec:PSNRRMSE}

In order to compare reconstructed depth images against original uncompressed data in Sec.~\ref{sec:DepthResults}, we rely on two common image quality metrics: {\emph{root-mean-square error}} (RMSE) and {\emph{peak signal-to-noise ratio}} (PSNR). We will introduce them here in the context of image comparison. Let $X$ be an $m\times n$ image and let $Y$ be an approximation of $X.$ RMSE is defined as its name suggests: specifically, it is the square root of the average of the set of squared differences in pixel values. That is, 
\begin{equation*}
\textrm{RMSE} := \sqrt{\frac{1}{mn}\sum_{i=1}^m\sum_{j=1}^n\left(X_{ij}-Y_{ij}\right)^2}.
\end{equation*}
Note that a consequence of this definition is that large differences in pixel values are exaggerated in their contributions to RMSE and small values ($<1$) have diminished contributions.
The definition of PSNR can be written in terms of RMSE, though interpretations of the two metrics differ. If $N$ is the maximum possible pixel value that can occur in $X,$ then we define PSNR for the approximation $Y$ to be 
\begin{equation*}
\textrm{PSNR} := 20\log_{10}\left(\frac{N}{\textrm{RMSE}}\right).
\end{equation*}

There are many image quality metrics~\cite{dosselmann2005existing}; we choose to focus on two of the most widely-used for the purpose of this work.

\subsection{Chemical Detection: ACE, MPACE, Thresholds}
\label{sec:ACEBCandThresholds}

The results we present in Sec.~\ref{sec:HSIResults} for reconstructed hyperspectral data involve videos of chemical releases. Thus our measure of quality is with respect to quantitative comparisons of chemical detection on reconstructed vs. raw data. We therefore present background on several relevant chemical detection techniques here.

A standard approach to chemical detection is via the {\emph{adaptive coherence/cosine estimator}} (ACE) algorithm~\cite{scharf1996adaptive,kraut2001adaptive}. Given a signature of a target chemical $s$ and a signature in a pixel $x$ in a hyperspectral cube, the ACE algorithm calculates the square of the cosine of the angle between $s$ and $x$ relative to background data. Specifically, if $\Gamma$ is the maximum likelihood estimator for the covariance matrix of the background, then the ACE statistic is 
$$\frac{(s^T\Gamma^{-1}x)^2}{(s^T\Gamma^{-1}s)(x^T\Gamma^{-1}x)}.$$

The signal can be strengthened in this setting by additionally applying a technique called {\emph{bulk coherence}} or {\emph{multipulse coherence}} (MPACE)~\cite{pakrooh2017adaptive,pakrooh2017adaptiveb,scharf2017multipulse}. This statistic is intended to enhance the signal in neighborhoods in which several pixels have high ACE values. Let $c_i$ be the value of the ACE statistic in pixel $i$ and consider a neighborhood of $M$ pixels. The bulk coherence statistic is then defined as 
$$1-\prod_{i=1}^M(1-c_i).$$
If several of the ACE values in the neighborhood are high (close to 1), then the product of the $(1-c_i)$ terms will be close to zero, resulting in a bulk coherence value close to 1. We notice significant improvement of signal strength as a result of the incorporation of bulk coherence. We further add noise reduction by including what we refer to as {\emph{persistence}}. Specifically, we add a filter that associates a value of zero to a pixel if the bulk coherence value in that pixel doesn't remain above a chosen threshold for five consecutive time frames. 

To make comparison of results objective, we algorithmically defined a threshold for determining whether a chemical is present or absent in a pixel based on the ACE statistic \cite{farnell2019TVvsL1}. We construct this threshold to be responsive to the device in use, the chemical of interest, and the reconstruction approach (e.g. $\ell_1$ optimization). Experimentally, this threshold appears to faithfully detect the presence or absence of chemicals in hyperspectral cubes and makes objective comparison possible.  




\section{DATA-DRIVEN SAMPLING}
\label{sec:data-driven}
As described in Sec.~\ref{sec:RIP}, the classical approach to sampling within the CS framework is to use a random sampling set. In this section we will present an alternative data-driven method. One of several potential benefits of our method is that it is flexible enough to accommodate hardware constraints that might make the use of traditional random sampling difficult. For this paper, we consider a setting in which hardware constraints specifically require either the use of entries of $\pm 1$ only or the use of entries of $0$ and $1$ only (the latter is relevant to the single pixel camera architecture). A natural choice in this context is to use the sampling basis provided by a Walsh-Hadamard transform or its `shifted' version. From the matrix for the transform, we select a subset of rows in order to perform CS at the desired level. Thus when we refer to sampling, we often refer to an ordering - the ordering of the rows determines the choice of subset since we simply select the first $k$ rows for sampling. The shifted version of a Walsh-Hadamard transform simply replaces $-1$ entries in the corresponding matrix with $0$ entries. Walsh-Hadamard transforms were defined as early as 1867 by Sylvester; they can be defined in various ways and the corresponding matrices have been given different names and orderings \cite{sylvester1867lx,fino1976unified,beauchamp1984applications,beer1981walsh}. We consider the standard ordering (`Hadamard order') of the rows of a Walsh-Hadamard matrix, the sequency ordering, an analog of sequency that we term frequency, random ordering, and the maximal-variance order as defined in terms of a relevant dataset. We also consider the corresponding orderings of the shifted Walsh-Hadamard matrices.

We briefly review some properties of Walsh-Hadamard matrices along with some important orderings of the rows of the Walsh-Hadamard matrix transform. Walsh-Hadamard matrices are square $(\pm 1)$-matrices and satisfy $W_mW_m^T=mI_m,$ where $W_m$ is an $m\times m$ Walsh-Hadamard matrix and $I_m$ is the $m\times m$ identity. We provide examples to illustrate various row orderings here. Let $m=8.$ Then the standard order produces
$$\begin{bmatrix*}[r]
1 & 1 & 1 & 1 & 1 & 1 & 1 & 1\\
1 & -1 & -1 & 1 & 1 & -1 & -1 & 1\\
1 & -1 & 1 & -1 & -1 & 1 & -1 & 1\\
1 & 1 & -1 & -1 & -1 & -1 & 1 & 1\\
1 & 1 & -1 & -1 & 1 & 1 & -1 & -1\\
1 & -1 & 1 & -1 & 1 & -1 & 1 & -1\\
1 & -1 & -1 & 1 & -1 & 1 & 1 & -1\\
1 & 1 & 1 & 1 & -1 & -1 & -1 & -1
\end{bmatrix*}.$$

The sequency order, in contrast, is defined so that the rows are in increasing order of zero-crossings between consecutive entries. Thus, the sequency order gives $$\begin{bmatrix*}[r]
1 & 1 & 1 & 1 & 1 & 1 & 1 & 1\\
1 & 1 & 1 & 1 & -1 & -1 & -1 & -1\\
1 & 1 & -1 & -1 & -1 & -1 & 1 & 1\\
1 & 1 & -1 & -1 & 1 & 1 & -1 & -1\\
1 & -1 & -1 & 1 & 1 & -1 & -1 & 1\\
1 & -1 & -1 & 1 & -1 & 1 & 1 & -1\\
1 & -1 & 1 & -1 & -1 & 1 & -1 & 1\\
1 & -1 & 1 & -1 & 1 & -1 & 1 & -1
\end{bmatrix*}.$$

Finally, inspired by the success of the use of randomness in standard CS settings, we will also compare against random orderings of the rows of Walsh-Hadamard matrices.

\subsection{Maximal-Variance Sampling}
In this section, we present the main definition of this paper: a data-driven method for selecting a subset of a basis for optimal sampling.

\noindent\textbf{Definition: Maximal-Variance Sampling} Choose a parameter $\alpha \in [0,1]$ that is the desired percent of compression. Let $X\in\mathbb{R}^{m\times n}$ be a set of training data of $n$ samples of points in $\mathbb{R}^m.$ Given a set of sampling vectors $S=\{s_i\}_{i=1}^p,$ with $p \geq \lceil (1-\alpha)m \rceil$, where $\lceil \cdot\rceil$ is the ceiling function, we compute the variance captured by each vector $s_i.$ Let $\sigma_i^2$ be the variance of the set $\{s_i\cdot x_j\}_{j=1}^n;$ equivalently $\sigma_i^2$ is the variance of the vector $s_i^TX.$ Define a permutation $\rho$ of $(1,2, \dots, p)$ so that $(\sigma_{\rho^{-1}(1)}^2, \sigma_{\rho^{-1}(2)}^2, \dots, \sigma_{\rho^{-1}(p)}^2)$ is decreasing. Then we define the maximal-variance sampling subset $M$ of $S$ to be 
\begin{equation*}
M=\{s_{\rho^{-1}(i)}\}_{i=1}^{\lceil(1-\alpha)m\rceil}.
\end{equation*}


The maximal-variance sampling algorithm proposed in the definition above is independent of the size of the sampling set. In the case where $S$ is defined by the Walsh-Hadamard matrix, the set of sampling vectors forms a basis for $\mathbb{R}^m,$ so $p=m.$ In general, the set $S$ need not have this property. The only requirement is that $p \geq \lceil (1-\alpha)m\rceil.$ 

\subsection{Frequency Ordering}
\label{sec:frequency}

While we propose maximal-variance sampling and demonstrate in Secs.~\ref{sec:DepthResults} and~\ref{sec:HSIResults} that it outperforms other orderings of rows from the Walsh-Hadamard matrix in terms of accuracy of reconstruction, it is sometimes preferable to have an ordering that does not require training data. Thus we also propose the following Walsh-Hadamard specific ordering, which is data independent and experimentally yields similarly strong results (see Secs.~\ref{sec:DepthResults} and~\ref{sec:HSIResults}). To sample scenes at a resolution of $m_1\times m_2$, where $m = m_1m_2$ is a power of $2$, we define an order on the rows of the Walsh-Hadamard matrix $W_m$ that we call the {\emph{frequency order}}. The frequency order is so called because it is a generalization of the sequency order that incorporates the 2-dimensional nature of the sampling. In the frequency order, the rows of the Walsh-Hadamard matrix are ordered in increasing order based on the number of connected components when each row is reshaped (via column-major format) to form an $m_1\times m_2$ matrix. For example, when $m=16,$ the frequency-ordered Walsh-Hadamard matrix has as its first row the vector of all ones; the second row is 
\begin{equation*}
\left[1, 1, -1, -1, 1, 1, -1, -1, 1, 1, -1, -1, 1, 1, -1, -1\right]
\end{equation*}
since it reshapes to a matrix with two connected components. Similarly, the remaining rows are ordered to produce increasing numbers of connected components in matrix form; the last row results in 16 components and is $\left[ 1, 0, 1, 0, 0, 1, 0, 1, 1, 0, 1, 0, 0, 1, 0, 1\right].$ Unlike the sequency order, the frequency order is not unique; for example, in this case two different rows have two connected components. When there is a tie for the number of connected components, we order the tied rows arbitrarily.

As we demonstrate in Sec.~\ref{sec:Performance}, the frequency order produces good performance consistently on various datasets. It is worth noting, however, that the frequency order is specific to the Walsh-Hadamard (or shifted Walsh-Hadamard) matrix because reshaped vectors naturally have meaningful connected components. If one wished to design a frequency ordering for a sampling matrix other than Walsh-Hadamard, it would likely be necessary to do thresholding or binning in order to produce a similar ordering. Even so, it remains to be determined whether such analogs of the frequency ordering are effective for sampling and reconstruction. In contrast, the maximal-variance ordering generalizes in a natural way for other sampling matrices.


\subsection{Performance on Data}
\label{sec:Performance}
We demonstrate the effectiveness of maximal-variance sampling on several datasets of interest. In each case, we simulate CS in software and compare accuracy of reconstructed data against the raw data. We begin in Sec.~\ref{sec:DepthResults} with results on a set of LIDAR depth images. 
\subsubsection{Depth imaging}
\label{sec:DepthResults}
In this section, we present results that compare maximal-variance sampling against other sampling orders of Walsh-Hadamard matrices on a dataset of over 4000 depth images collected with a Swiss Ranger SR4000 Time of Flight camera at Colorado State University. Each depth image has a spatial resolution of $64\times 64.$

\begin{figure}[!htb]
\begin{center}
\begin{tabular}{c}
\includegraphics[height=5cm]{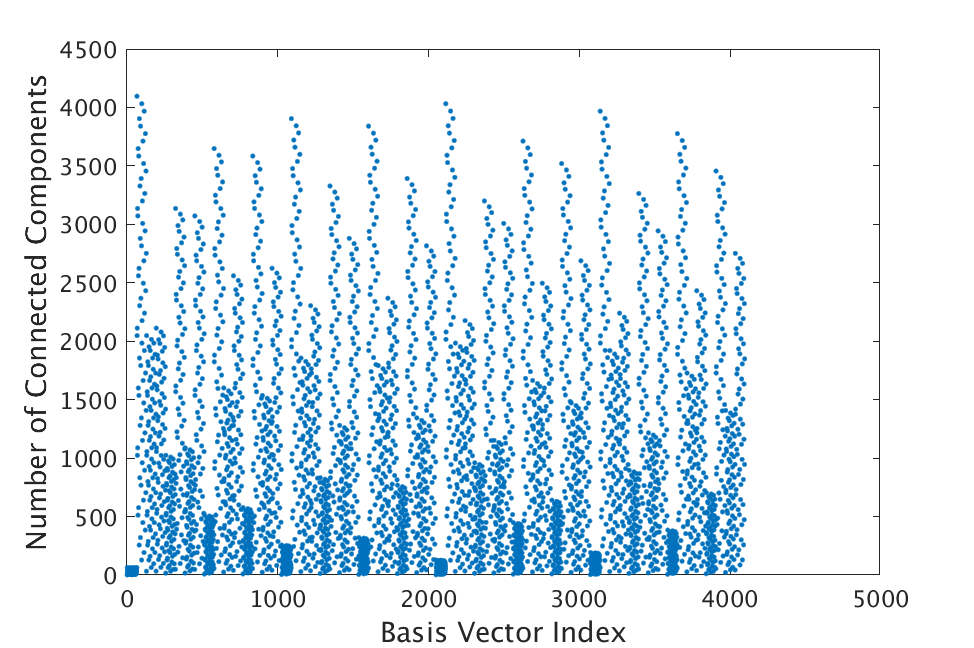}\\
\includegraphics[height=5cm]{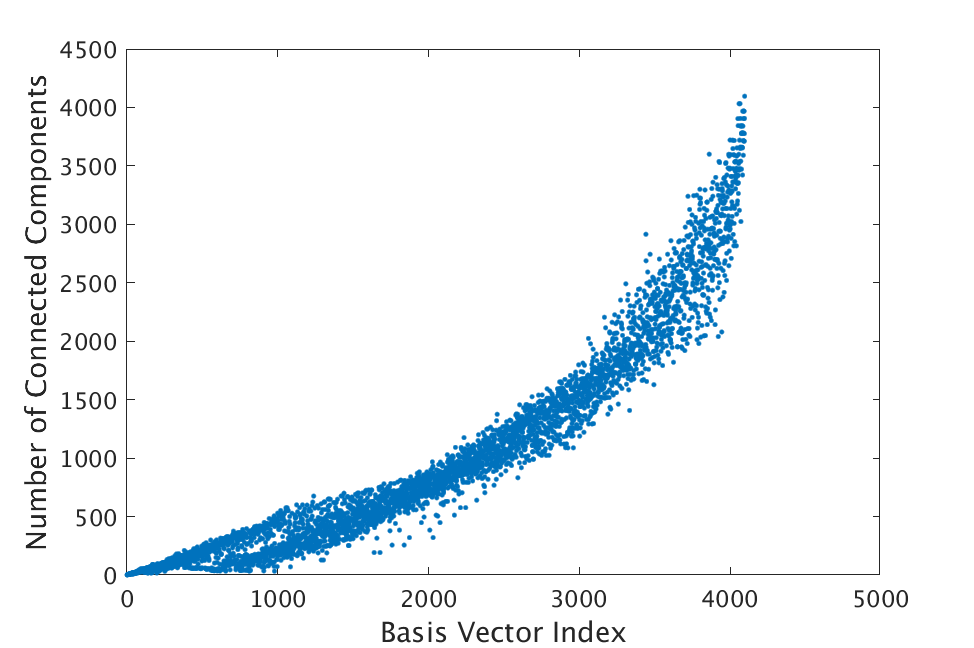}\\
\includegraphics[height=5cm]{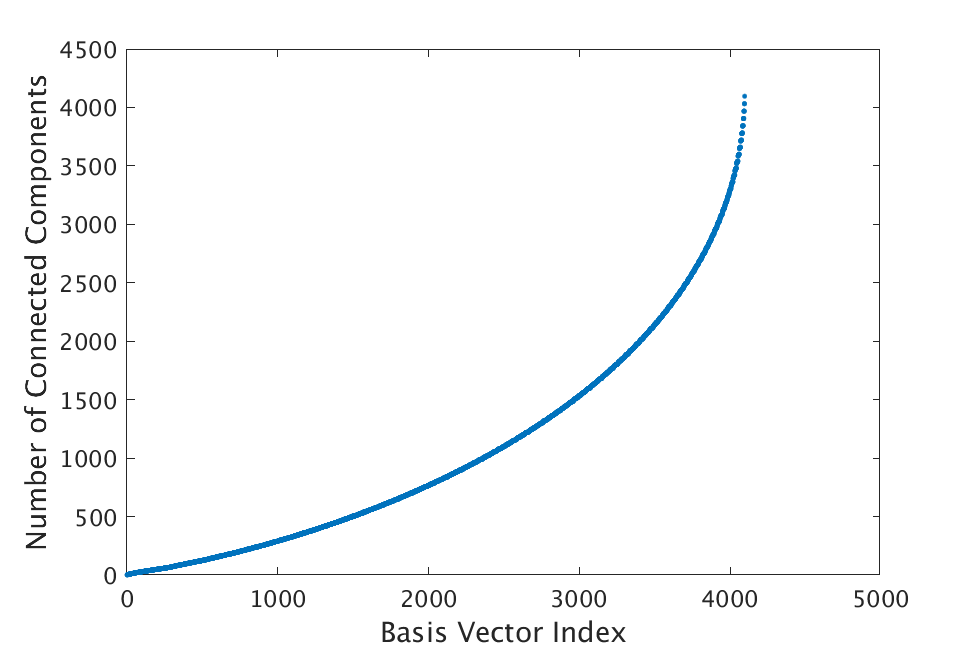}
\end{tabular}
\end{center}
\caption{\label{fig:frequency}} Comparison of the number of connected components in reshaped sampling vectors (drawn from rows of a $4096 \times 4096$ Walsh-Hadamard matrix) for various orders. Top to bottom: standard order, maximal-variance order for the Swiss Ranger dataset, and frequency order.
\end{figure}

In Fig.~\ref{fig:frequency}, we show the number of connected components in the sampling basis vectors corresponding to various orders: standard, maximal-variance for the full Swiss Ranger data, and frequency. Note that the maximal-variance order defined for the Swiss Ranger depth data seems in general to preference sampling vectors with lower numbers of connected components. 

In the remainder of this section, we split the Swiss Ranger depth dataset into a training and testing set. The training set was taken to be the first 80\% of the images and the testing set was taken to be the remaining 20\%. Note that we intentionally split the data in this way as opposed to taking a random sample of 80\%. The data is collected sequentially as the camera moves around a scene, so data points that appear within a small time range are generally similar to each other. Thus a random sample would actually provide an unfair advantage in training.

\begin{figure}[!htb]
\begin{center}
\includegraphics[height=9cm]{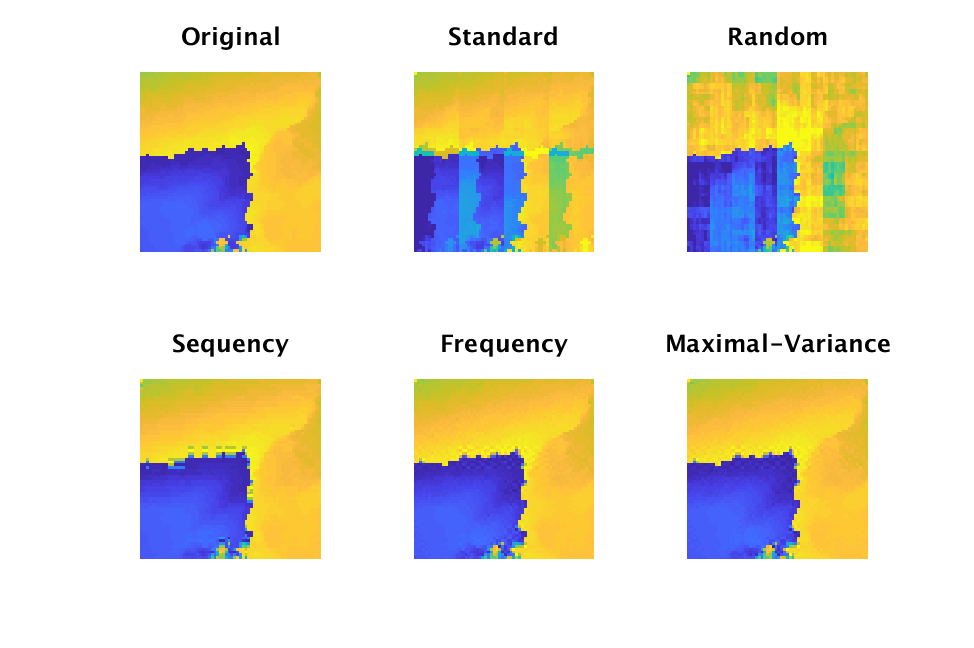}
\end{center}
\caption{\label{fig:SRreconstructions}} Example reconstructions of a test image (top left) sampled by various orders of rows of the shifted Walsh-Hadamard matrix. The reconstructed images labeled by sampling basis (with corresponding PSNR) are: standard ordering (49.4), random ordering (49.6), sequency order (58.0), frequency order (62.7), and  maximal-variance order (62.8). Note that the standard and random sampling orders produce artifacts in the reconstructions. All simulated sampling is at $25\%$ compression on $64\times 64$ depth images. All reconstructions use split Bregman iteration with the Haar wavelet sparsity-promoting basis.
\end{figure}

We compare reconstructions after simulated CS against an original test set depth image in Fig. ~\ref{fig:SRreconstructions}. Note that the maximal-variance and frequency orders perform best. The corresponding PSNR values are: standard order (49.4), random order (49.6), sequency order (58.0), frequency order (62.7), and maximal-variance order (62.8).

\begin{figure}[!htb]
\begin{center}
\includegraphics[height=9cm]{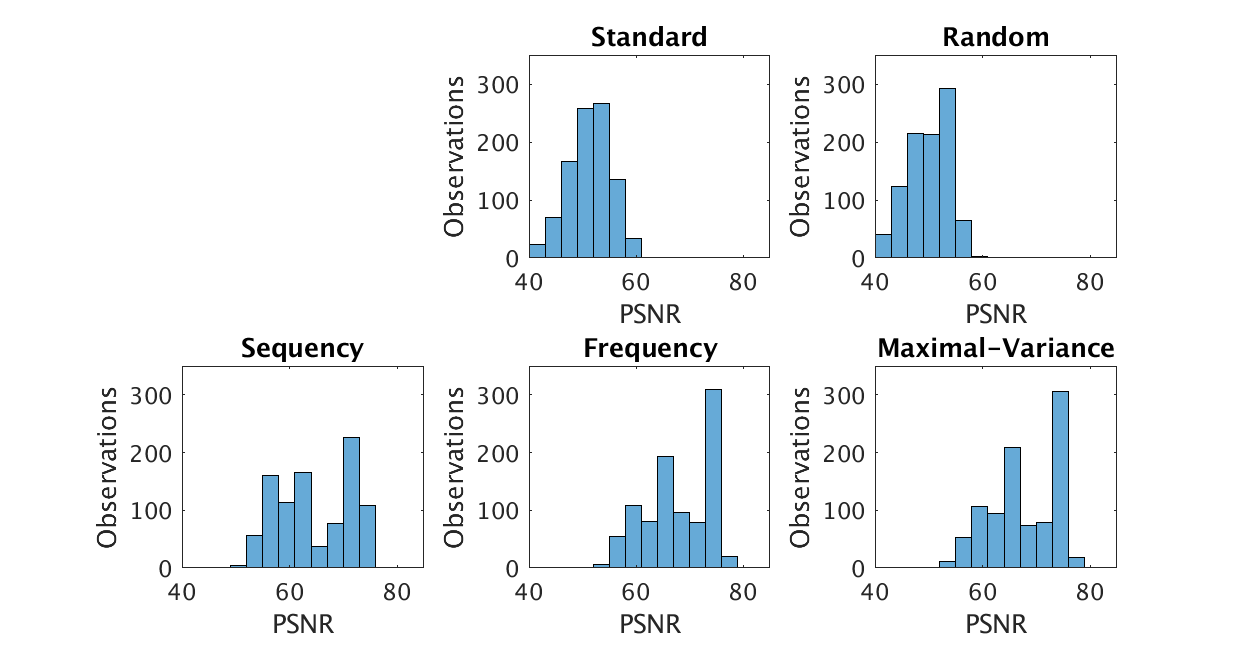}
\end{center}
\caption{\label{fig:PSNRhist}}
A comparison of the quality of reconstruction of depth images from the Swiss Ranger test set after using different sampling orders from the shifted Walsh-Hadamard matrix. Each subplot is a histogram of the PSNR values for reconstructed images sampled with that particular order. The organization of the subplots coincides with that in Fig. \ref{fig:SRreconstructions}. All sampling is at $25\%$ compression on $64\times 64$ depth images. All reconstructions use split Bregman iteration with a Haar wavelet sparsity-promoting basis. 
\end{figure}

\begin{figure}[!htb]
\begin{center}
\begin{tabular}{c}
\includegraphics[height=5.5cm]{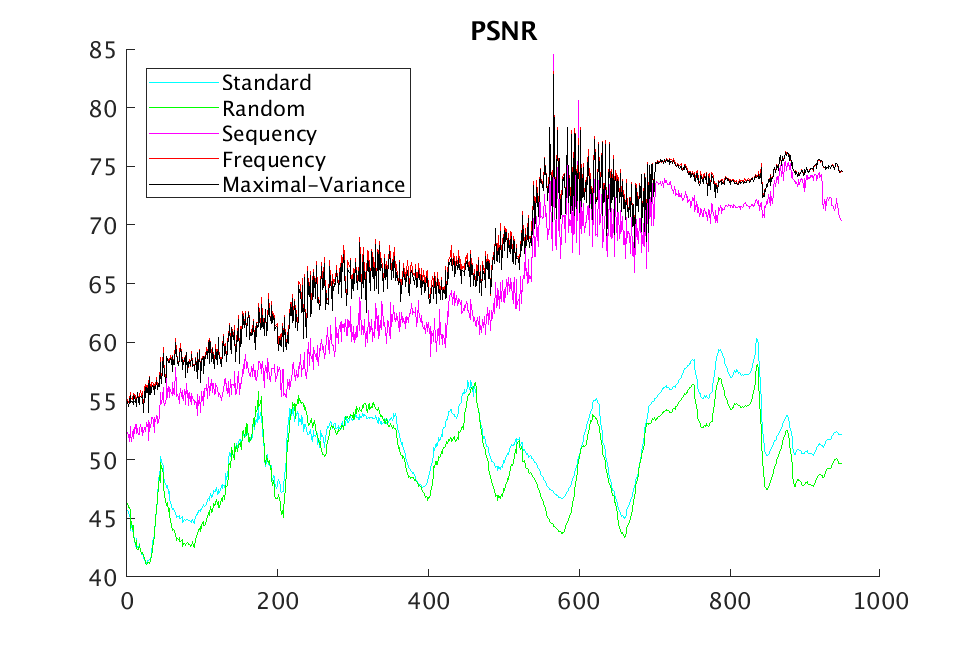}
\includegraphics[height=5.5cm]{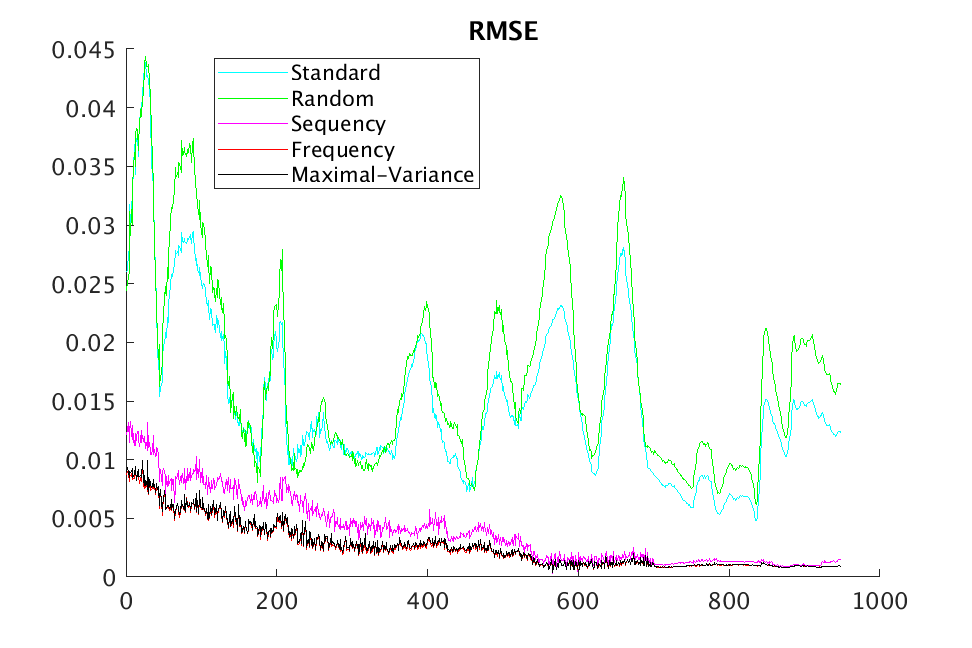}

\end{tabular}
\end{center}
\caption{\label{fig:PSNRRMSE}}
A comparison of the quality of reconstruction of depth images from the Swiss Ranger test set after using different sampling orders from the shifted Walsh-Hadamard matrix. Left: PSNR for each test image. Right: RMSE for each test image. All sampling was done at 25\% compression on $64\times 64$ depth images. All reconstructions use split Bregman iteration with a Haar wavelet sparsity-promoting basis. 
\end{figure}

In Figs.~\ref{fig:PSNRhist} and~\ref{fig:PSNRRMSE}, we provide quantitative comparisons of the performance of sampling and reconstruction with the various sampling orders on the Swiss Ranger depth image test set. Fig.~\ref{fig:PSNRhist} provides a histogram of PSNR values for each of the five sampling orders on the test set. Note that frequency and maximal-variance orders have higher occurrences of relatively large PSNR values. Fig.~\ref{fig:PSNRRMSE} provides a comparison of PSNR and RMSE on each test image for each of the different sampling orders. Again we see the best performance arising from frequency and maximal-variance sampling. The mean PSNR value across the test set for the maximal-variance order is 67.7, whereas that for standard and random orders is 51.3 and 50.1, representing a 32\% and 35\% increase, respectively. The mean PSNR value for the frequency order is similar to that for the maximal-variance order, demonstrating that both the maximal-variance order and the frequency order produce strong results on this particular dataset.

\subsubsection{Hyperspectral imaging for chemical detection}
\label{sec:HSIResults}
In this section, we present results for multi/hyper-spectral datasets that contain chemical explosions. The two datasets we consider are the Physical Sciences Inc.  Fabry-P\'{e}rot interferometer sensor multispectral dataset and the Johns Hopkins Applied Physics Lab FTIR-based longwave infrared sensor hyperspectral dataset. We present analysis of detection of a chemical release in each dataset: a GAA-release in Fabry-P\'{e}rot and a release of SF6 in Johns Hopkins. We select a $64\times 64$ spatial subset of the cubes and include all associated spectral bands. In both datasets, the impact of sampling is apparent in the results. Chemical detection is the main motivation for this particular CS application; thus our performance comparison emphasizes detection over other potential metrics like PSNR.

\begin{figure}[ht]
\begin{center}
\begin{tabular}{c}
\includegraphics[height=3.6cm]{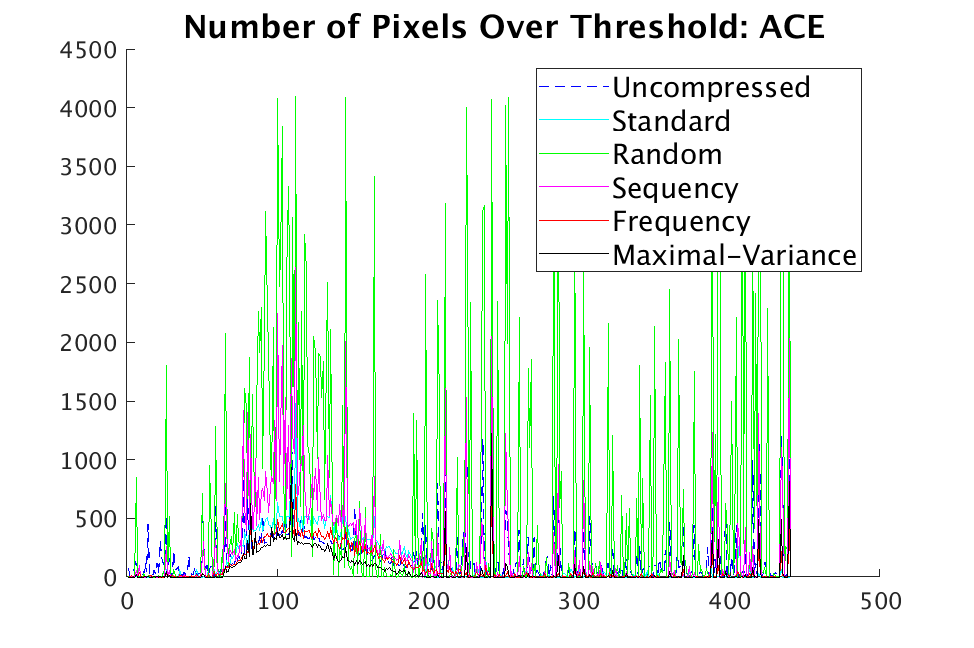}
\includegraphics[height=3.6cm]{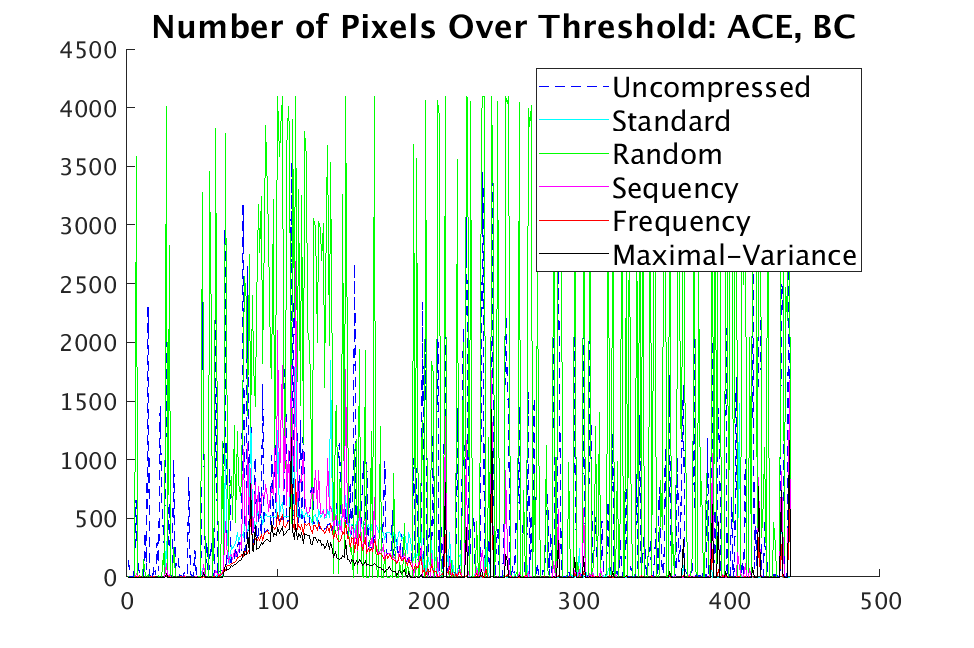}
\includegraphics[height=3.6cm]{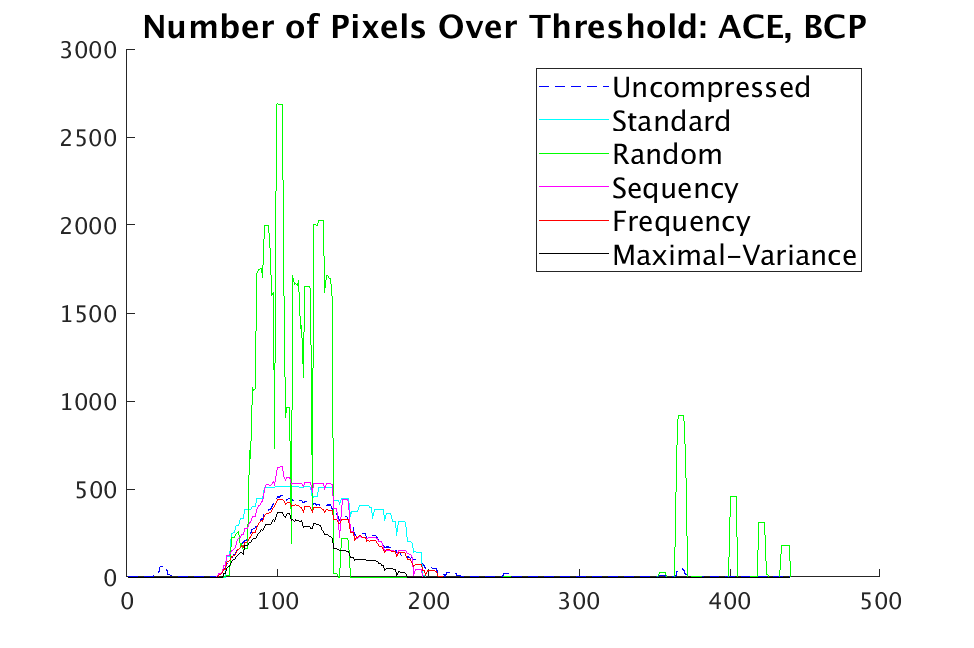}
\end{tabular}
\end{center}
\caption{\label{fig:GAAnumberover}} Comparison of performance of various sampling orderings on Fabry-P\'{e}rot GAA data. Each plot shows the number of pixels over the threshold as a function of the frame in the hyperspectral video. From left to right, the plots show detection by: ACE, ACE with bulk coherence, and ACE with bulk coherence and persistence.
\end{figure}

In Fig.~\ref{fig:GAAnumberover}, we provide a plot of the number of pixels over the associated threshold (defined based on ACE values in background cubes - see Sec.~\ref{sec:ACEBCandThresholds}). The three subfigures show the different results for different variants of the ACE chemical detection algorithm. Moving from left to right, the subfigures show the number of pixels over the threshold for ACE, ACE with bulk coherence, and ACE with bulk coherence and persistence. The rightmost subfigure (which includes the most post-processing) is the easiest to visually examine given the large amount of noise in this particular video.

\begin{figure}[!htb]
\begin{center}
\begin{tabular}{c}
\includegraphics[height=5.5cm]{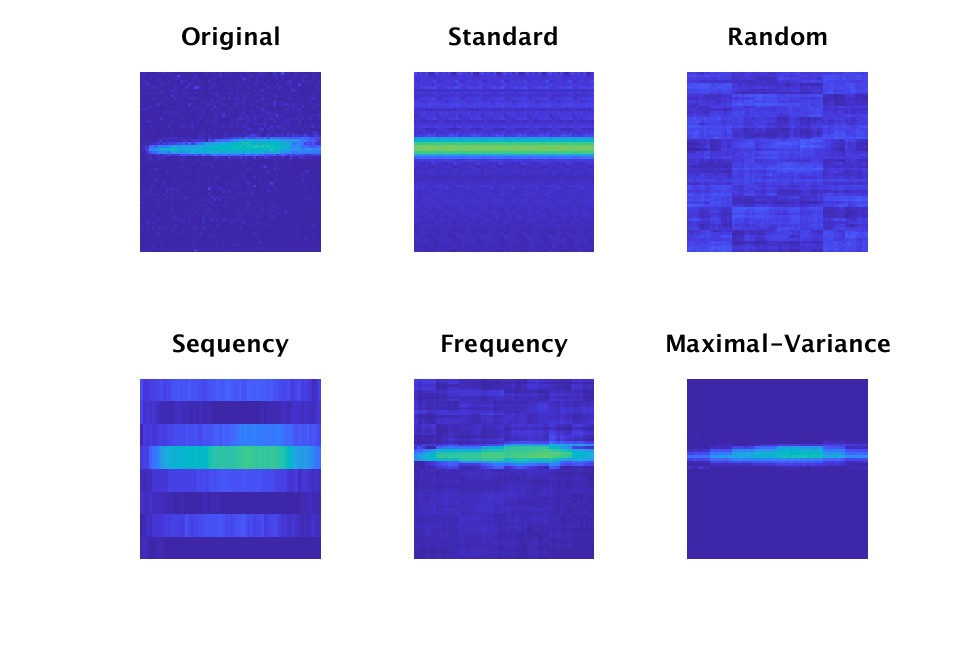}
\includegraphics[height=5.5cm]{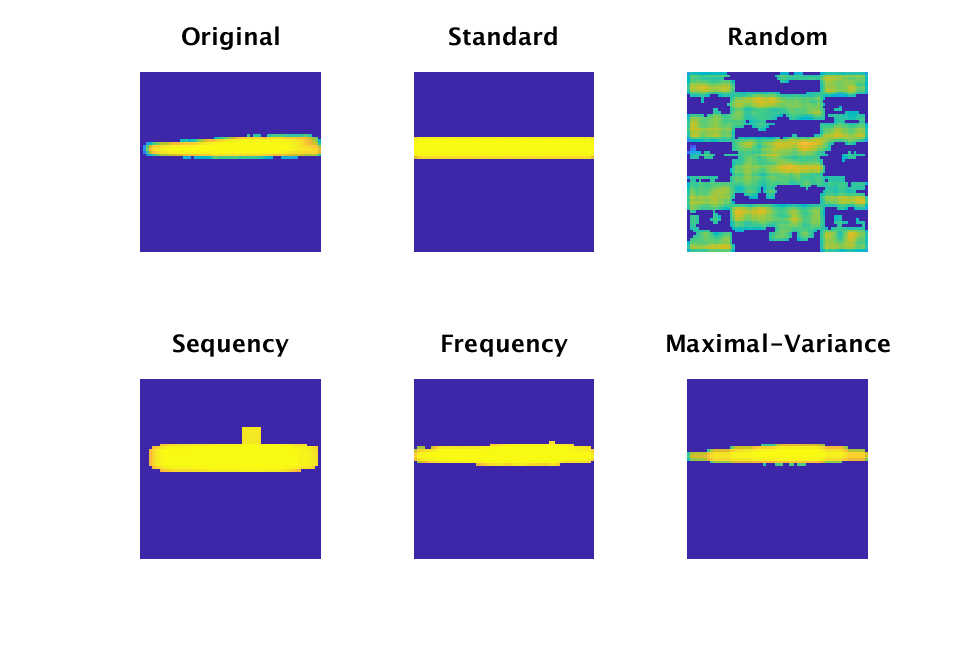}
\end{tabular}
\end{center}
\caption{\label{fig:GAAexreconstructionACE}} Left: ACE on an example reconstruction from Fabry-P\'{e}rot GAA (Cube 100). Right: ACE with bulk coherence and persistence on the same reconstructed cube. Note that unwanted artifacts occur in the reconstructions after CS with standard, random, and sequency orders. Furthermore, the corresponding reconstructions make it difficult or impossible to localize the presence of the chemical simulant, especially in the raw ACE values.
\end{figure}

It is apparent from  Fig.~\ref{fig:GAAnumberover} that the standard, random, and sequency orders produce artificially high numbers of pixels with ACE values over the threshold. The reason for this is made clear from an examination of an example reconstruction. In Fig.~\ref{fig:GAAexreconstructionACE}, we apply ACE to reconstructions of Fabry-P\'{e}rot GAA cube 100 after sampling with various orders. Note that the standard, random, and sequency orders have artifacts and make it impossible to localize the signal. The signal appears strongest in the reconstruction based on the frequency order, but the localization is strongest with the maximal-variance order.

We next consider a release of SF6 in the Johns Hopkins hyperspectral dataset (specifically the SF6 27 Romeo dataset). In this analysis, we include both the maximal-variance order defined on Johns Hopkins SF6 data and that defined on Fabry-P\'{e}rot GAA data to emphasize the robustness of the orders to different settings. Even though the chemical and device are both different, the GAA maximal-variance order performs well in the SF6 context. 

We chose the SF6 dataset because of the dynamic behavior of the chemical within the $64\times 64$ spatial field of view we selected. First, the chemical is released, then it effectively disappears from the scene, and then it reappears (presumably due to a quick change in wind direction). This made for an interesting performance test - specifically, we were curious about which orderings would lead to successful detection of the chemical in the time frames when the chemical reappears, necessarily in a weaker, dissipated form. 

\begin{figure}[ht]
\begin{center}
\begin{tabular}{c}
\includegraphics[height=3.6cm]{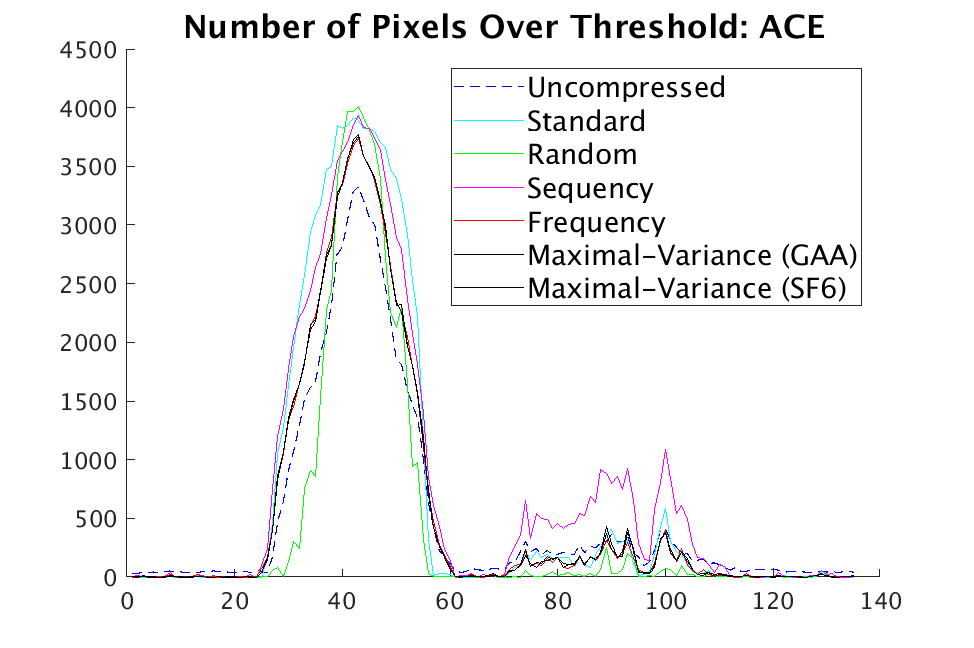}
\includegraphics[height=3.6cm]{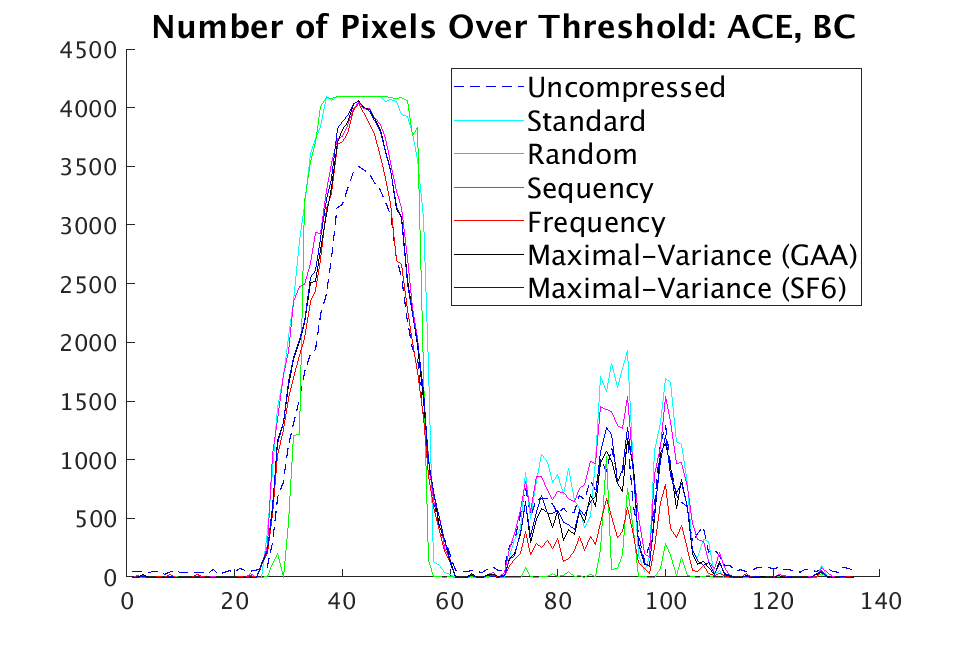}
\includegraphics[height=3.6cm]{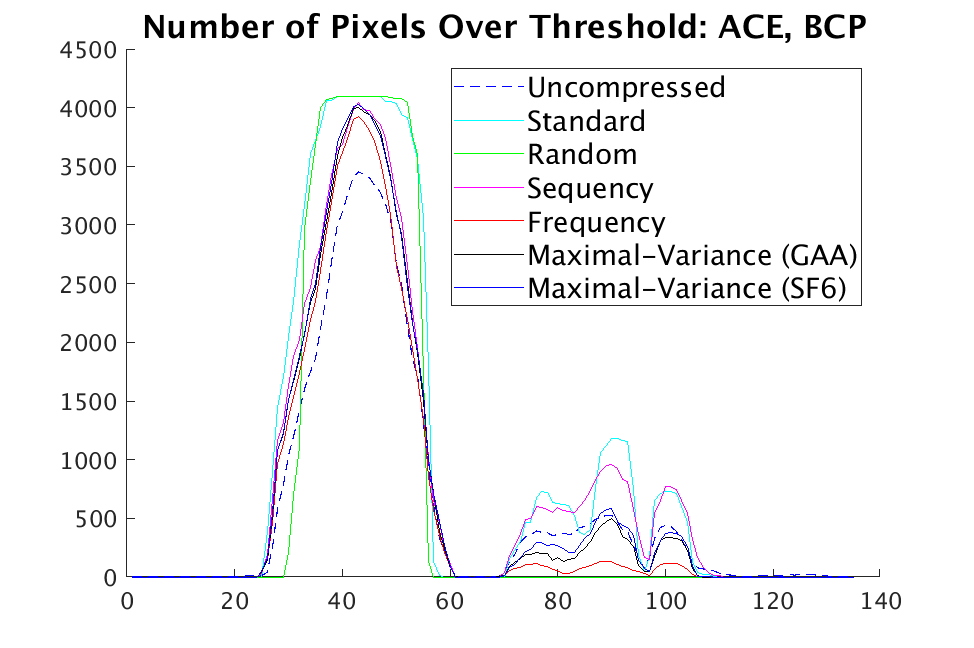}
\end{tabular}
\end{center}
\caption{\label{fig:SF6numberover}} Comparison of performance of various sampling orderings on Johns Hopkins SF6 data. Each plot shows the number of pixels over the threshold as a function of the frame in the hyperspectral video. From left to right, the plots show detection by: ACE, ACE with bulk coherence, and ACE with bulk coherence and persistence. While every ordering results in chemical detection that matches that seen in the uncompressed data (dashed blue curve), from example reconstructions (Figs.~\ref{fig:SF6exreconstructionACE30} and~\ref{fig:SF6exreconstructionACE90}) it is clear that the standard, random, and sequency orders have artifacts and often fail to localize the signal.
\end{figure}

\begin{figure}[ht]
\begin{center}
\begin{tabular}{c}
\includegraphics[height=5.5cm]{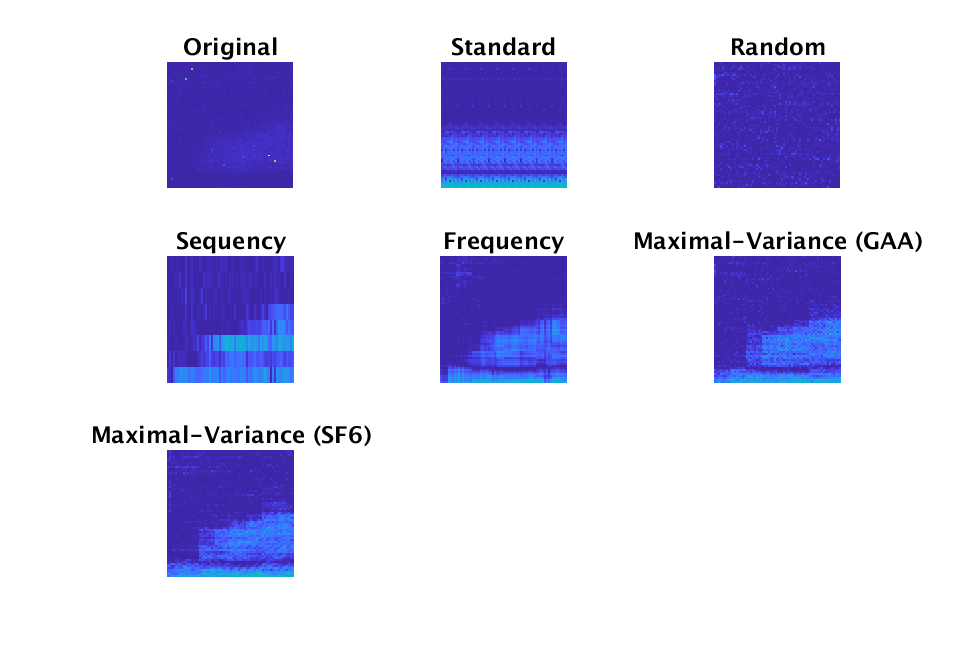}
\includegraphics[height=5.5cm]{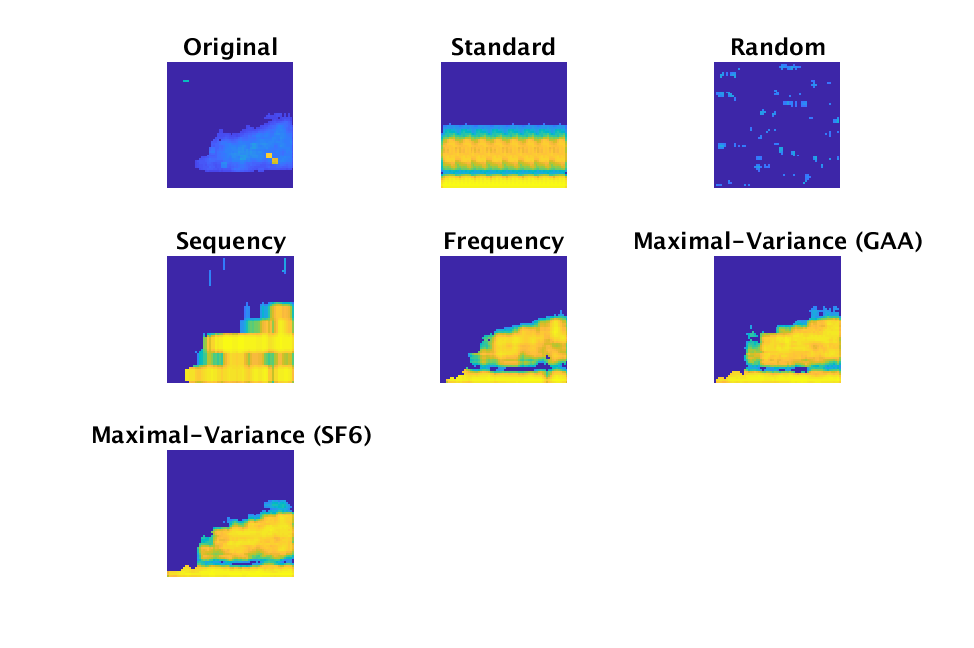}
\end{tabular}
\end{center}
\caption{\label{fig:SF6exreconstructionACE30}} Left: ACE on an example reconstruction from Johns Hopkins SF6 data (Cube 30). Right: ACE, bulk coherence, and persistence on the same reconstructed cube. Unwanted artifacts occur in the reconstructions after CS with standard, random, and sequency orders, and these orders make it impossible to localize the presence of the chemical in the reconstruction. Note that all figures are displayed on a range of $[0,1]$ with the exception of the original, uncompressed data, which are displayed on ranges of $[0,0.05]$ and $[0,0.07],$ respectively for the left and right figures. These ranges for uncompressed data are used because the ACE values are sufficiently low as to make the chemical invisible otherwise. 
\end{figure}

\begin{figure}[ht]
\begin{center}
\begin{tabular}{c}
\includegraphics[height=5.5cm]{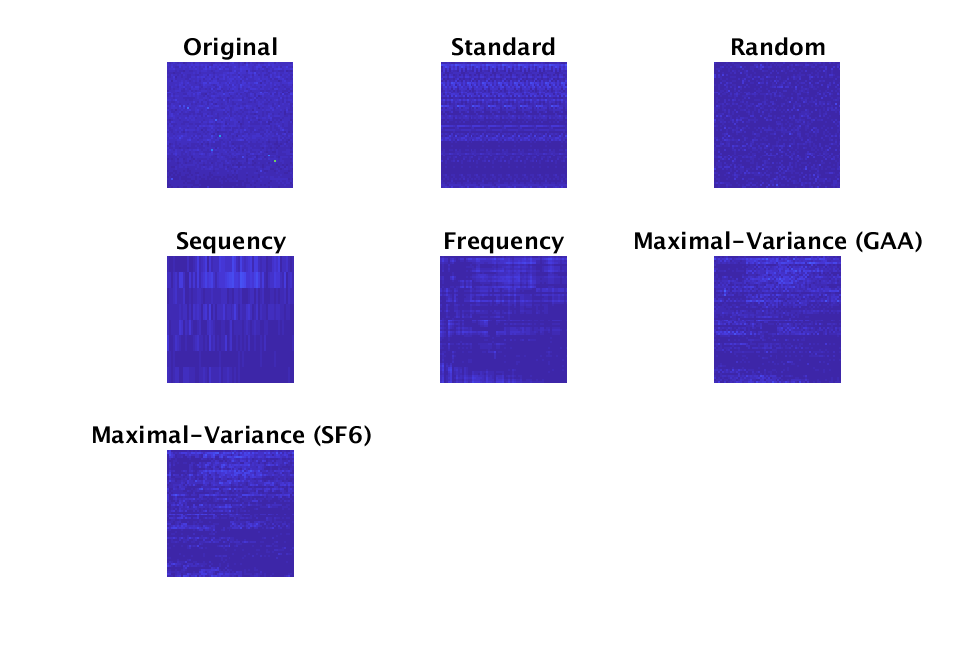}
\includegraphics[height=5.5cm]{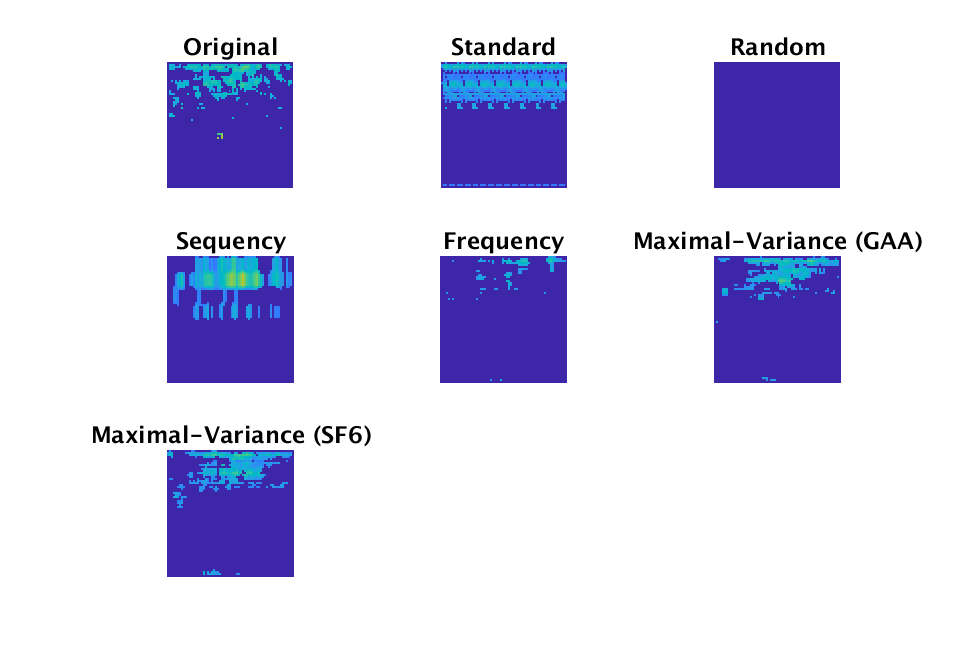}
\end{tabular}
\end{center}
\caption{\label{fig:SF6exreconstructionACE90}} Left: ACE on an example reconstruction from Johns Hopkins SF6 data (Cube 90). Right: ACE, bulk coherence, and persistence applied to the same reconstructed cubes. Unwanted artifacts occur in the reconstructions sampled with standard, random, and sequency orders, and these orders make it impossible to localize the presence of the chemical in the reconstruction. Note that all figures are displayed on a range of $[0,1]$ with the exception of the original, uncompressed ACE data and the original, uncompressed ACE data with bulk coherence and persistence, which are both displayed on a range of $[0,0.02]$. This range is used for the uncompressed data because the ACE values are sufficiently low as to make the chemical invisible otherwise. 
\end{figure}

In Fig.~\ref{fig:SF6numberover}, we show the number of pixels over the threshold after applying ACE, ACE with bulk coherence, and ACE with bulk coherence and persistence to the reconstructed cubes that were sampled with various orderings of the shifted Walsh-Hadamard basis. As in the GAA example, it is the case that some orderings detect the chemical in a high number of pixels and consequently appear to be producing good performance. However, from example reconstructions (e.g. those in Figs.~\ref{fig:SF6exreconstructionACE30} and~\ref{fig:SF6exreconstructionACE90}), it is apparent that the standard, random, and sequency orders produce reconstructions that fail to localize the signal.

From Figs.~\ref{fig:SF6exreconstructionACE30} and~\ref{fig:SF6exreconstructionACE90}, we observe that the orderings that produce reconstructions that both detect the presence of SF6 and do a reasonable job of localizing that presence are frequency, maximal-variance (from GAA data), and maximal-variance (from SF6 data). Combining this with the information displayed in Fig.~\ref{fig:SF6numberover}, we observe that during the time when the chemical reappears in the scene, the SF6 maximal-variance order does best, then the GAA maximal-variance order, then frequency. (Recall that while standard, random, and sequency have many pixels above the threshold, the reconstructions themselves are not informative.)

\subsubsection{Example: maximal-variance order sampling vectors}
\label{sec:MVOexamples}
In this section, we provide example sampling basis vectors in the order defined by maximal-variance for the Swiss Ranger depth data. Note that there is a clear preference in the basis ordering for large-scale features.

\begin{figure}[ht]
\begin{center}
\begin{tabular}{c}
\includegraphics[height=5cm]{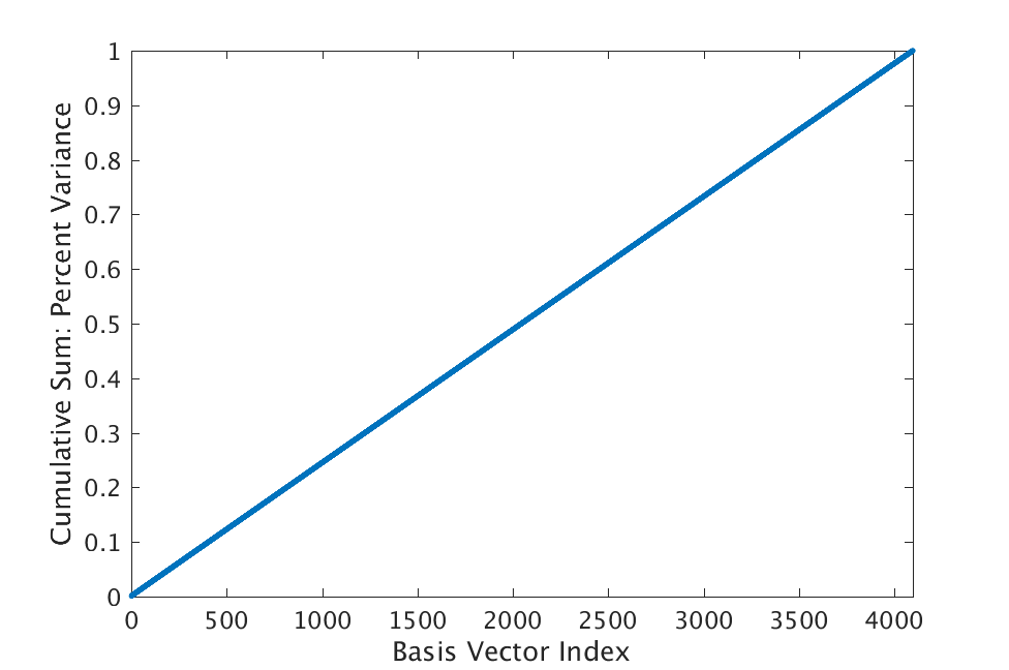}
\includegraphics[height=5cm]{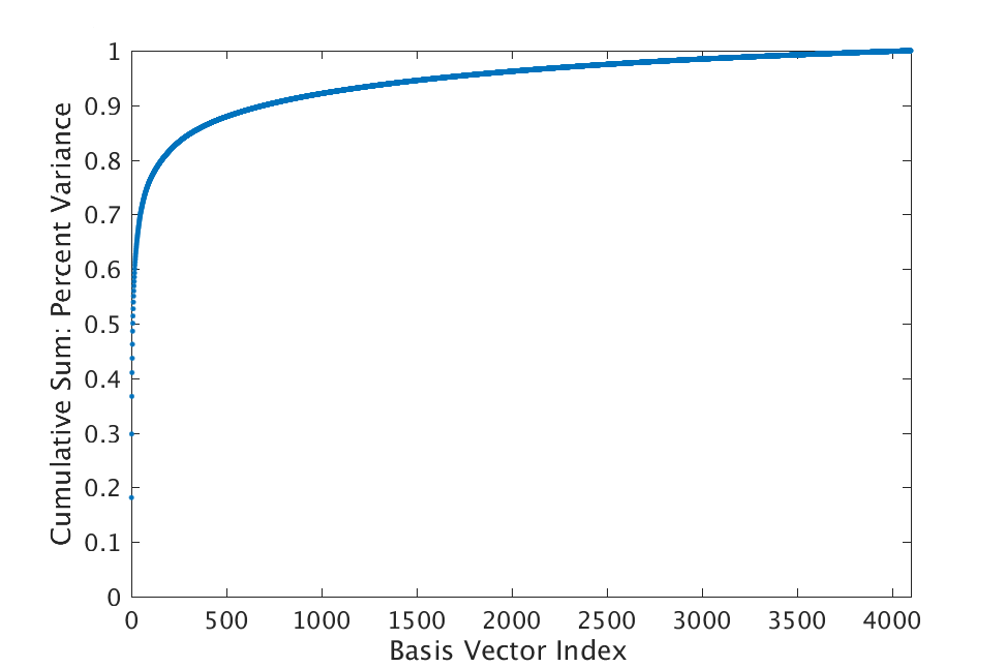}
\end{tabular}
\end{center}
\caption{\label{fig:cumulativevar}} Comparison of cumulative percent variance explained by successive inclusion of ordered sampling basis vectors in maximal-variance order for data that has and has not been mean-subtracted. Left: Variance explained by sampling vectors on original Swiss-Ranger data. Right: Variance explained by sampling vectors on mean-subtracted Swiss-Ranger data.
\end{figure}

We note here that the maximal-variance ordering for the shifted Walsh-Hadamard matrix is more effective if the data is mean-subtracted prior to defining the ordering. To be precise, we subtract the mean of each data point from itself individually; this is somewhat different than what is often meant by mean-subtraction for datasets. In Fig.~\ref{fig:cumulativevar}, we show the cumulative percent of variance explained by the ordered shifted Walsh-Hadamard basis for the two cases in which the data has and has not been mean-subtracted. Note that we see a significant improvement in how quickly variance is captured in the case in which data is mean-subtracted. It is also worth pointing out here that it is possible to incorporate this form of mean-subtraction into CS by using the sample corresponding to a vector of all $1$'s. 

Mean-subtraction has little effect on the maximal-variance ordering for the Walsh-Hadamard matrix. Additionally, in our explorations, we have seen quite similar orderings produced for the Walsh-Hadamard matrix and the shifted Walsh-Hadamard matrix on mean-subtracted data.

\begin{figure}[ht]
\begin{center}
\begin{tabular}{c}
\includegraphics[height=5.5cm]{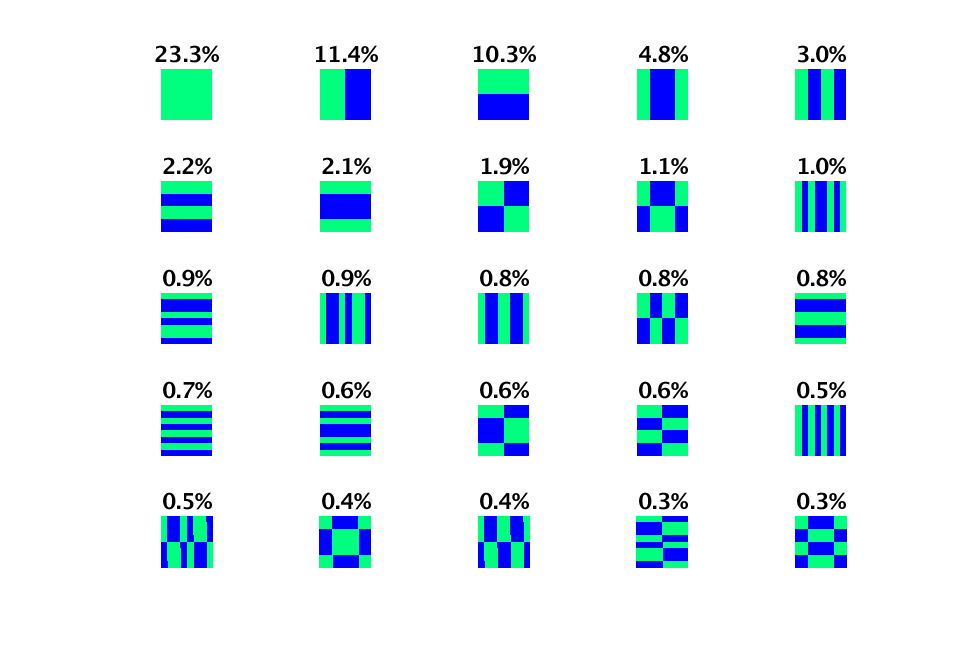}
\includegraphics[height=5.5cm]{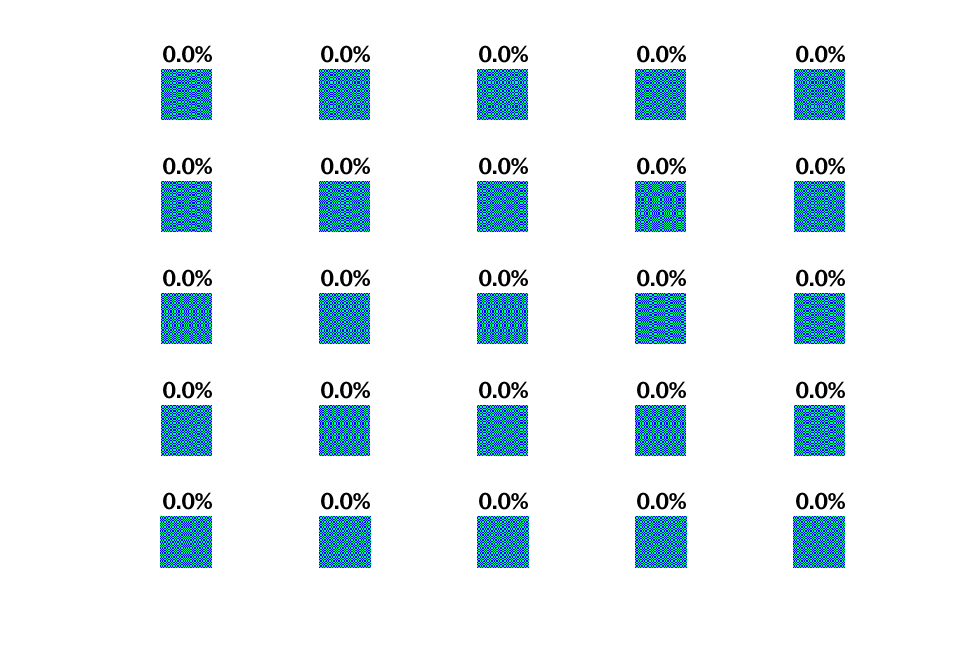}
\end{tabular}
\end{center}
\caption{\label{fig:optimalbasisvecs}} Examples of Walsh-Hadamard sampling basis vectors that appear at the beginning and at the end of the maximal-variance order for the Swiss Ranger depth training data, along with corresponding approximate percent variance. Left: First 25 sampling basis vectors. Right: Last 25 sampling basis vectors. In this case, we trained the order using the Walsh-Hadamard matrix rather than the shifted Walsh-Hadamard matrix.
\end{figure}

In Fig.~\ref{fig:optimalbasisvecs}, we show examples of (re-shaped) sampling vectors that appear early in the order and late in the order to provide some context for the nature of the maximal-variance order. Recall that the vectors are ordered by variance of the corresponding set of dot products of samples. The associated percent of the total variance across all sampling vectors is shown above each reshaped sampling vector. Note that the sampling vectors that show up early appear to capture relatively large-scale features and those that show up late appear to generally capture high frequency information.

\section{CONCLUSION}
\label{sec:conclusion}.

In this paper, we presented a method for selecting CS sampling vectors from a sampling basis in a manner that is responsive to training data. Specifically, we defined the maximal-variance sampling order and we demonstrated the benefits of this method of sampling in the context of both depth images and multi/hyper-spectral images. We further proposed and compared results for a `frequency' order that is appropriate for sampling bases such as Walsh-Hadamard bases and requires no training. These sampling methods are particularly valuable in settings with hardware constraints and settings in which context-specific sampling is desired.


We suggest the following questions as directions for further study:
\begin{itemize}
    \item It would be interesting to study methods in which the choice of sampling matrix described in this paper is continually adapted on the fly as new data is acquired. Will this adaptive form of sampling produce better results in applications in which we receive a continuous stream of data?
    \item In this paper we built $S$ starting from a Walsh-Hadamard matrix whose entries had been shifted to only be $0$'s and $1$'s. It would be interesting to explore how this method performs when $S$ is initialized with another matrix.
    \item The method proposed in this paper was only tested with respect to a single optimization problem and a single sparsity-promoting basis. In the future we would like to expand our study to include explorations of how our method performs when other optimization problems are used (for example, minimizing total variation or using custom dictionaries as sparsity-promoting bases). 
\end{itemize}

\acknowledgments 
The authors would like to thank Louis Scharf for insightful discussions related to this work, especially with regard to content involving ACE and MPACE. This research was partially supported by 
Department of Defense Army STTR Compressive Sensing Flash IR 3D Imager contract W911NF-16-C-0107
and Department of Energy STTR Compressive Spectral Video in the LWIR contract W911SR-17-C-0012.

\bibliography{report} 

\def\cprime{$'$} \def\cprime{$'$} \def\cprime{$'$} \def\cprime{$'$}
  \def\cprime{$'$} \def\cprime{$'$} \def\cprime{$'$}
\begin{thebibliography}{10}

\bibitem{duarte2008single}
Duarte, M.~F., Davenport, M.~A., Takhar, D., Laska, J.~N., Sun, T., Kelly,
  K.~F., and Baraniuk, R.~G., ``Single-pixel imaging via compressive
  sampling,'' {\em IEEE signal processing magazine}~{\bf 25}(2),  83--91
  (2008).

\bibitem{lustig2007sparse}
Lustig, M., Donoho, D., and Pauly, J.~M., ``Sparse {MRI}: The application of
  compressed sensing for rapid {MR} imaging,'' {\em Magnetic Resonance in
  Medicine: An Official Journal of the International Society for Magnetic
  Resonance in Medicine}~{\bf 58}(6),  1182--1195 (2007).

\bibitem{candes2005decoding}
Candes, E.~J. and Tao, T., ``Decoding by linear programming,'' {\em IEEE
  transactions on information theory}~{\bf 51}(12),  4203--4215 (2005).

\bibitem{donoho2001uncertainty}
Donoho, D.~L. and Huo, X., ``Uncertainty principles and ideal atomic
  decomposition,'' {\em IEEE transactions on information theory}~{\bf 47}(7),
  2845--2862 (2001).

\bibitem{takhar2006new}
Takhar, D., Laska, J.~N., Wakin, M.~B., Duarte, M.~F., Baron, D., Sarvotham,
  S., Kelly, K.~F., and Baraniuk, R.~G., ``A new compressive imaging camera
  architecture using optical-domain compression,'' in [{\em Computational
  Imaging IV}{\nolinebreak\hspace{0.1em}]},   {\bf 6065},  606509,
  International Society for Optics and Photonics (2006).

\bibitem{baraniuk2007compressive}
Baraniuk, R.~G., ``Compressive sensing [lecture notes],'' {\em IEEE signal
  processing magazine}~{\bf 24}(4),  118--121 (2007).

\bibitem{baraniuk2010model}
Baraniuk, R.~G., Cevher, V., Duarte, M.~F., and Hegde, C., ``Model-based
  compressive sensing,'' {\em IEEE Transactions on Information Theory}~{\bf
  56}(4),  1982--2001 (2010).

\bibitem{cosofret2009airis}
Cosofret, B.~R., Chang, S., Finson, M.~L., Gittins, C.~M., Janov, T.~E., Konno,
  D., Marinelli, W.~J., Levreault, M.~J., and Miyashiro, R.~K., ``{AIRIS}
  standoff multispectral sensor,'' in [{\em Chemical, Biological, Radiological,
  Nuclear, and Explosives (CBRNE) Sensing X}{\nolinebreak\hspace{0.1em}]},
  {\bf 7304},  73040Y, International Society for Optics and Photonics (2009).

\bibitem{broadwater2011primer}
Broadwater, J.~B., Limsui, D., and Carr, A.~K., ``A primer for chemical plume
  detection using {LWIR} sensors,'' {\em Technical Paper, National Security
  Technology Department, Las Vegas, NV}  (2011).

\bibitem{willett2014sparsity}
Willett, R.~M., Duarte, M.~F., Davenport, M.~A., and Baraniuk, R.~G.,
  ``Sparsity and structure in hyperspectral imaging: Sensing, reconstruction,
  and target detection,'' {\em IEEE signal processing magazine}~{\bf 31}(1),
  116--126 (2014).

\bibitem{GO09}
Goldstein, T. and Osher, S., ``The split {B}regman method for
  {$L1$}-regularized problems,'' {\em SIAM J. Imaging Sci.}~{\bf 2}(2),
  323--343 (2009).

\bibitem{candes2008restricted}
Candes, E.~J., ``The restricted isometry property and its implications for
  compressed sensing,'' {\em Comptes rendus mathematique}~{\bf 346}(9-10),
  589--592 (2008).

\bibitem{candes2006robust}
Cand{\`e}s, E.~J., Romberg, J., and Tao, T., ``Robust uncertainty principles:
  Exact signal reconstruction from highly incomplete frequency information,''
  {\em IEEE Transactions on information theory}~{\bf 52}(2),  489--509 (2006).

\bibitem{donoho2006compressed}
Donoho, D.~L., ``Compressed sensing,'' {\em IEEE Transactions on information
  theory}~{\bf 52}(4),  1289--1306 (2006).

\bibitem{baraniukjohnson}
Baraniuk, R., Davenport, M., DeVore, R., and Wakin, M., ``The
  {J}ohnson-{L}indenstrauss lemma meets compressed sensing. 2006,'' {\em
  Preprint} .

\bibitem{dosselmann2005existing}
Dosselmann, R. and Yang, X.~D., ``Existing and emerging image quality
  metrics,'' in [{\em Electrical and Computer Engineering, 2005. Canadian
  Conference on}{\nolinebreak\hspace{0.1em}]},   1906--1913, IEEE (2005).

\bibitem{scharf1996adaptive}
Scharf, L.~L. and McWhorter, L.~T., ``Adaptive matched subspace detectors and
  adaptive coherence estimators,'' in [{\em Signals, Systems and Computers,
  1996. Conference Record of the Thirtieth Asilomar Conference
  on}{\nolinebreak\hspace{0.1em}]},   1114--1117, IEEE (1996).

\bibitem{kraut2001adaptive}
Kraut, S., Scharf, L.~L., and McWhorter, L.~T., ``Adaptive subspace
  detectors,'' {\em IEEE Transactions on signal processing}~{\bf 49}(1),  1--16
  (2001).

\bibitem{pakrooh2017adaptive}
Pakrooh, P., Scharf, L.~L., Cheney, M., Homan, A., and Ferrara, M.,
  ``Multipulse adaptive coherence for detection in wind turbine clutter,'' {\em
  IEEE Transactions on Aerospace and Electronic Systems}~{\bf 53}(6),
  3091--3103 (2017).

\bibitem{pakrooh2017adaptiveb}
Pakrooh, P., Scharf, L., Cheney, M., Homan, A., and Ferrara, M., ``The adaptive
  coherence estimator for detection in wind turbine clutter,'' in [{\em Radar
  Conference (RadarConf), 2017 IEEE}{\nolinebreak\hspace{0.1em}]},
  1793--1798, IEEE (2017).

\bibitem{scharf2017multipulse}
Scharf, L.~L. and Pakrooh, P., ``Multipulse subspace detectors,'' in [{\em
  Conference record-Asilomar Conference on Signals, Systems, \&
  Computers}{\nolinebreak\hspace{0.1em}]},  (2017).

\bibitem{farnell2019TVvsL1}
Farnell, E., Kvinge, H., Dupuis, J.~R., Kirby, M., Peterson, C., and Schundler,
  E.~C., ``Total variation vs {L1} regularization: a comparison of compressive
  sensing optimization methods for chemical detection,'' (2019).
\newblock Preprint.

\bibitem{sylvester1867lx}
Sylvester, J.~J., ``{LX}. thoughts on inverse orthogonal matrices, simultaneous
  signsuccessions, and tessellated pavements in two or more colours, with
  applications to {N}ewton's rule, ornamental tile-work, and the theory of
  numbers,'' {\em The London, Edinburgh, and Dublin Philosophical Magazine and
  Journal of Science}~{\bf 34}(232),  461--475 (1867).

\bibitem{fino1976unified}
Fino, B.~J. and Algazi, V.~R., ``Unified matrix treatment of the fast
  {W}alsh-{H}adamard transform,'' {\em IEEE Transactions on Computers} (11),
  1142--1146 (1976).

\bibitem{beauchamp1984applications}
Beauchamp, K.~G.,  [{\em Applications of {W}alsh and related functions: with an
  introduction to sequency theory}{\nolinebreak\hspace{0.1em}]}, Academic press
  (1984).

\bibitem{beer1981walsh}
Beer, T., ``Walsh transforms,'' {\em American Journal of Physics}~{\bf 49}(5),
  466--472 (1981).

\end{thebibliography}
\bibliographystyle{spiebib} 

\end{document}